\documentstyle[11pt,aas2pp4]{article}

\slugcomment{Accepted for publication in ApJ Letters)}

\lefthead{Zapatero Osorio et al.}
\righthead{An L-type substellar object in Orion}

\begin{document}

\title{An L-type substellar object in Orion: reaching the mass boundary 
       between brown dwarfs and giant planets}

\author{M. R. Zapatero Osorio, V. J. S. B\'ejar, R. Rebolo\altaffilmark{1}}
\affil{Instituto de Astrof\'\i{}sica de Canarias, 
       E--38200 La Laguna, Tenerife, Spain}

\author{E. L. Mart\'\i n, and G. Basri}
\affil{Astronomy Department, University of California, Berkeley, CA 94720, 
       USA}

\altaffiltext{1}{Consejo Superior de Investigaciones Cient\'\i ficas, CSIC. 
                 Spain}

\centerline{{\sl e-mail addresses}: mosorio@ll.iac.es, vbejar@ll.iac.es, 
            rrl@ll.iac.es}
\centerline{ege@popsicle.berkeley.edu, basri@soleil.berkeley.edu}

\begin{abstract}
We present near-infrared photometry ($J$-band) and low-resolution optical 
spectroscopy (600--1000\,nm) for one of the faintest substellar member 
candidates in the young $\sigma$\,Orionis cluster, S\,Ori\,47 ($I=20.53$, 
B\'ejar et al. \cite{bejar99}). Its very red ($I-J$)\,=\,3.3$\pm$0.1 color 
and its optical spectrum allow 
us to classify S\,Ori\,47 as an L1.5-type object which fits the 
low-luminosity end of the cluster photometric and spectroscopic sequences. 
It also displays atmospheric features indicative of low gravity such as 
weak alkaline lines and hydride and oxide bands, consistent with the 
expectation for a very young object still undergoing gravitational collapse. 
Our data lead us to conclude that S\,Ori\,47 is a true substellar member of 
the $\sigma$\,Orionis cluster. Additionally, we present the detection of 
Li\,{\sc i} in its atmosphere which provides an independent confirmation 
of youth and substellarity. Using current theoretical evolutionary 
tracks and adopting an age interval of 1--5\,Myr for the $\sigma$\,Orionis 
cluster, we estimate the mass of S\,Ori\,47 at 0.015$\pm$0.005\,$M_\odot$, 
i.e. at the minimum mass for deuterium burning, which has been proposed as 
a definition for the boundary between brown dwarfs and giant planets. 
S\,Ori\,47 could well be the result of a natural extension of the process 
of cloud fragmentation down to the deuterium burning mass limit; a less 
likely alternative is that it has originated from a protoplanetary disc 
around a more massive cluster member and later ejected from its orbit due 
to interacting effects within this rather sparse ($\sim$12 objects pc$^{-3}$) 
young cluster. The study of this object serves as a guide for future deep 
searches for free-floating objects with planetary masses.
\end{abstract}

\keywords{open clusters and associations: individual ($\sigma$\,Orionis) --- 
          stars: evolution --- stars: low-mass, brown dwarfs}

\section{Introduction}

Nearby, young clusters offer unique opportunities to establish the 
observational properties of objects at and below the hydrogen burning 
limit (brown dwarfs, BDs), and constitute an ideal scenario to deepen 
in our knowledge of the initial mass function (IMF) in the substellar 
regime. The advantage of searching in these clusters is that important 
parameters like age, distance, and metallicity are relatively well known, 
and thus, they can be used for constraining masses and luminosities of 
cluster members. The first claims of the discovery of BDs in the Pleiades 
(Rebolo, Zapatero Osorio \& Mart\'\i n \cite{rebolo95}; Basri, Marcy \& 
Graham \cite{basri96}) have been followed by recent deeper searches which 
considerably extend the area surveyed in this cluster (Bouvier et al. 
\cite{bouvier98}; Stauffer et al. \cite{stauffer98}; Zapatero Osorio et 
al. \cite{osorio99}) as well as in other young nearby clusters like 
Praesepe (Magazz\`u et al. \cite{magazzu98}), Hyades (Gizis, Reid \& 
Monet \cite{gizis99}), $\rho$\,Oph (Comeron et al. \cite{comeron98}), 
Chamaleon~I (Comeron, Rieke, \& Neuh\"auser \cite{comeron99}), and IC\,348 
(Luhman et al. \cite{luhman98}, \cite{luhman99}). These efforts were 
devoted to determine how far the IMF extends into the substellar domain. 
To date, BDs with masses of only 0.035\,$M_\odot$ have been confirmed in 
the Pleiades (Mart\'\i n et al. \cite{martin98}), and down to roughly 
0.025\,$M_\odot$ in a few younger clusters (Luhman, Liebert \& Rieke 
\cite{luhman97}; Mart\'\i n, Basri \& Zapatero Osorio \cite{martin99a}; 
B\'ejar, Zapatero Osorio \& Rebolo \cite{bejar99}). 

Very recently, B\'ejar et al. (\cite{bejar99}; hereafter referred to as 
BOR99) have performed a deep $RIZ$ survey in the young $\sigma$\,Orionis 
cluster (Walter et al. \cite{walter97}), located within the Orion 1b 
association. This work reveals a rich population of very low-mass stars 
and BDs in the cluster with short ages in the interval 1--5\,Myr. About 
40 substellar candidate members were found down to $I$\,=\,21~mag. 
We present here the follow-up IR photometry and optical 
spectroscopy of a $\sigma$\,Orionis faint candidate, S\,Ori\,47, which was 
selected for its very red $I-Z$ color. We have estimated a mass of 
0.015\,$M_\odot$ for this object, confirming it as the least massive 
free-floating object discovered to date in the Orion star-forming region. 

\section{Observations and Analysis}

Near-infrared observations ($J$-band) of the substellar candidate S\,Ori\,47 
were performed in 1998 Dec. 12 and in 1999 Feb. 24 at the 1.5\,m 
Carlos S\'anchez Telescope (TCS, Teide Observatory on the Island of 
Tenerife) using the CIR infrared camera equipped with an HgCdTe 
256$\times$256 array. This detector provided 1\arcsec ~pixels and a field 
of view of about 4\arcmin$\times$4\arcmin. The total integration time per 
night was 1080\,s, the final image being the co-addition of twelve dithered 
exposures of 90\,s each. Raw data were processed using standard techniques 
within the IRAF\footnote {IRAF is distributed by National Optical Astronomy 
Observatory, which is operated by the Association of Universities for 
Research in Astronomy, Inc., under contract with the National Science 
Foundation.} environment. Instrumental aperture magnitudes were transformed 
into the UKIRT system by observing faint standard stars (Casali 
\& Hawarden \cite{casali92}) at different airmasses. The photometric 
calibration of both nights showed a dispersion of $\pm$0.2\,mag. We can 
give a smaller error bar to the measurement of S\,Ori\,47 because close 
to it (at $\sim$\,42\arcsec E, 10\arcsec S) and within the field of view of 
our images there is another cluster BD candidate (S\,Ori\,27, M7 
spectral type, BOR99) for which there is IR data of high precision (B\'ejar 
et al. \cite{bejar00}); S\,Ori\,27 does not appear to be variable. We have 
also performed differential photometry of our target with respect to this 
brighter object. Table~\ref{tab1} summarizes the optical and the averaged 
$J$ photometry of S\,Ori\,47 and Fig.~\ref{fig1} shows its location in 
the $I$ versus $I-J$ color-magnitude diagram. 

Low-resolution spectroscopy was collected with the Keck II Low Resolution 
Imaging Spectrograph (Oke et al. \cite{oke95}) in 1998 Dec. 21 
using the longslit mode. At the beginning of the exposure we rotated the 
slit until parallactic angle in order to minimize light losses due to 
refraction. We used a slit width projection onto the detector of 4\,pixels 
(1\arcsec.2) and the 150\,g/mm grating blazed at 750\,nm, which gave a 
nominal dispersion of 4.8\,\AA/pix. One single exposure of 1800\,s was taken. 
The spectrum was debiased, flat-fielded, optimal extracted and wavelength 
calibrated using the emission spectra of HgNeAr lamps. All these procedures 
were performed within IRAF. The instrumental signature of the observed data 
was removed by matching the spectrum of BRI\,0021--0214 obtained with the 
same instrumental configuration during that night with a previous one that 
we already had flux calibrated (Mart\'\i n, Rebolo \& Zapatero Osorio 
\cite{martin96}). Figure~\ref{fig2} (upper panel) displays the final 
spectrum of S\,Ori\,47 (resolution $\sim$20\,\AA) where some features have 
been indicated as in Kirkpatrick et al. (\cite{kirk99}).

The colors and optical spectrum of S\,Ori\,47 are indicative 
of a cool object belonging to the recently proposed L spectral class 
(Mart\'\i n et al. \cite{martin97}). Molecular absorption band heads of 
TiO in the range 640--740\,nm appear very weak in our spectrum while the 
hydrides CaH, CrH and FeH are as strong as VO. We have used the 
classification schemes proposed by Kirkpatrick et al. (\cite{kirk99}) and 
Mart\'\i n et al. (\cite{martin99b}) based on several atomic, molecular 
and pseudocontinuum indices to determine an L1.5-type for S\,Ori\,47 with 
a dispersion of half a subclass. Its optical spectrum shows great 
similarities with other young BDs of similar spectral types 
like G\,196-3B (Rebolo et al. \cite{rebolo98}) and Roque\,25 (Mart\'\i n 
et al. \cite{martin98}). The strengths of the CrH and FeH bands and the 
Na\,{\sc i}, Cs\,{\sc i} and K\,{\sc i} lines are lower in S\,Ori\,47 
relative to other L1--L2 type objects in the field discovered by the all-sky 
IR survey Denis (Delfosse et al. \cite{delfosse97}; Mart\'\i n et al. 
\cite{martin99b}). We attribute this effect to low gravity of S\,Ori\,47 
demonstrating its youth. Its pre-main sequence nature is also inferred 
from the location of this object in Fig.~\ref{fig1}. S\,Ori\,47 clearly lies 
above the photometric sequence of field dwarfs and nicely fits the 
extrapolation toward low luminosities of the $\sigma$\,Orionis cluster 
optical/IR sequences. There are photometric and spectroscopic evidences 
for low reddening in the region where S\,Ori\,47 was found ($A_V\,\le\,0.1$, 
BOR99). We note that the pattern of the cluster members in 
the $I$ versus $I-J$ diagram is well reproduced by the models (Burrows et 
al. \cite{burrows97} and Baraffe et al. \cite{baraffe98}) overplotted to 
the observations. The NextGen models by Baraffe et al. provide magnitudes 
and colors directly; we have used their bolometric corrections for an age 
of 5\,Myr in order to convert luminosities given by Burrows et al. into 
magnitudes. The spectroscopy and photometry of S\,Ori\,47 support its 
membership in the $\sigma$\,Orionis cluster. Because of its cool temperature 
and low luminosity the substellarity of this object is thus guaranteed. 

An independent consistency check of the substellar nature of S\,Ori\,47 is 
given by the detection of lithium in its atmosphere. Figure~\ref{fig2} 
(lower panel) shows an enlargement around the Li\,{\sc i} 670.8\,nm line of 
the low-resolution spectrum of S\,Ori\,47 in comparison to two other L1--L2 
BDs in the field, Kelu\,1 (Ruiz, Leggett \& Allard \cite{ruiz97}) and 
G\,196-3B. The spectra were taken with identical instrumental configuration 
and during the same observing night. However, the signal-to-noise ratio 
of S\,Ori\,47, which is the faintest of the three objects, is lower 
at these wavelenghts and we can only claim a detection at 2--3\,$\sigma$ 
confidence level. The EW we measure is given in Table~\ref{tab1}. The 
presence of lithium in such a cool object is enough to confirm its 
substellar nature with a mass $\le$0.05\,$M_{\odot}$ (Baraffe et al. 
\cite{baraffe98}; D'Antona \& Mazzitelli \cite{dantona94}). However, 
the evidence of its cluster membership from the color-magnitude diagram 
and spectral features will provide a better constraint on the mass and 
luminosity of S\,Ori\,47. 

H$\alpha$ is not observed in strong emission (upper limit of 6\,\AA) at 
the resolution of S\,Ori\,47's spectrum in Fig.~\ref{fig2}. This is not 
unusual for young objects with very cool temperatures. Among field L-dwarfs 
only few cases are known with H$\alpha$ in emission, and typically EWs are 
below 4\,\AA ~(Tinney et al. \cite{tinney97}; Kirkpatrick et al. 
\cite{kirk99}; Mart\'\i n et al. \cite{martin99b}). The chromospheric 
activity in the $\sigma$\,Orionis cluster seems to diminish with decreasing 
mass along the cluster sequence. M-type BDs in the cluster show 
indications of considerable activity (BOR99); S\,Ori\,47 with a much later 
class does not present evidences of an active chromosphere. This behaviour 
is also observed among Pleiades very low-mass members (Zapatero Osorio et 
al. \cite{osorio97}; Mart\'\i n et al. \cite{martin98}). 

\section{Discussion and Final Remarks}

We have estimated the probability that S\,Ori\,47 could be a contaminating 
field object (and not a member of $\sigma$\,Orionis) using the number 
density (0.01\,pc$^{-3}$) of early-L type objects found in the field 
(Delfosse et al. \cite{delfosse99}; Reid et al. \cite{reid99}). Adopting 
an absolute M$_I$ magnitude of 15.1\,mag for an L1--L2-type dwarf and 
considering the completeness limit and covered area of BOR99's survey, we 
obtain an extremely low contamination ($<$1\%). The discovery of 
S\,Ori\,47 can be understood in terms of the much higher 
stellar/substellar density in the Orion region than in the field. On the 
basis of our optical/IR photometry and low-resolution spectroscopy we can 
safely conclude that S\,Ori\,47 is a member of the cluster. 

Several authors have addressed the temperature scale for L-dwarfs 
(Leggett, Allard \& Haus\-childt \cite{leggett98}; Basri et al. 
\cite{basri99}; Pavlenko, Zapatero Osorio \& Rebolo \cite{pavlenko99}), 
and they all suggest on a value of 2000$\pm$100\,K for an L2-type object. 
However, this is a calibration done for field dwarfs with gravities larger 
than that of S\,Ori\,47. Because the $T_{\rm eff}$ scale for cool, 
low-gravity objects is several hundred degrees warmer (Luhman et al. 
\cite{luhman97}), an upward correction of 100--200\,K may be required 
for S\,Ori\,47, which is consistent with the predictions by evolutionary 
models. The luminosity of this object can be obtained from the Hipparcos 
distance (352\,pc) to the star $\sigma$\,Orionis, which belongs to the 
cluster of the same name, and using the $J$ bolometric correction. We have 
estimated the latter from the Lyon ``dusty'' models (I. Baraffe 1999, 
private communication) as a function of the $I-J$ color, resulting 
BC$_J$\,=\,2.17. This yields a luminosity for S\,Ori\,47 of 
log\,$L/L_{\odot}$\,=\,--2.75$\pm$0.15\,dex. 

Figure~\ref{fig3} displays the luminosity evolution of objects of different 
masses according to the recent evolutionary models of Burrows et al. 
(\cite{burrows97}) and the Lyon ``dusty'' models . S\,Ori\,47 is indicated 
in this diagram with a box that takes into account the error bar in 
luminosity and the age interval adopted for the $\sigma$\,Orionis cluster 
(1--5\,Myr with a most probable age of 3\,Myr, BOR99). The mass of 
S\,Ori\,47 can be derived from this figure at 0.015$\pm$0.005\,$M_\odot$, 
where the error bar comes from uncertainties in age and other observable 
parameters. We note that the models used here and those by D'Antona \& 
Mazzitelli (\cite{dantona97}) predict very similar luminosities for young, 
low-mass objects, and therefore our result does not change significantly 
from model to model. However, we shall caution that the realibility of this 
mass estimate relies on the evolutionary tracks, and thus, it is subject 
to possible systematic effects arising from the description of 
atmospheres. Nevertheless, in relative terms we can safely conclude that 
S\,Ori\,47 is the least massive substellar object so far identified in 
the Orion region, and moreover, it is the least massive L-type substellar 
object for which photometry and spectroscpy have allowed a mass 
determination with some precision. 

The mass of S\,Ori\,47 is remarkably close to the deuterium burning mass 
limit (DBL, 0.013--0.015\,$M_{\odot}$, Saumon et al. \cite{saumon96}; 
D'Antona \& Mazzitelli \cite{dantona97}). It has 
been proposed recently that planets should be defined as those substellar 
objects that never fuse deuterium, while BDs do it efficiently for some 
portion of their evolution (see Oppenheimer, Kulkarni \& Stauffer 
\cite{oppenheimer99}, and references therein). The error bars in the mass 
determination of S\,Ori\,47 can effectively locate it on either side of 
the DBL, so at this stage we cannot argue whether it 
is a very low-mass BD which will eventually deplete its initial deuterium 
content or a free-floating massive, giant planet that will preserve 
deuterium for its whole life. S\,Ori\,47 provides evidence that objects 
with masses around the DBL can form in Nature. The fact that 
more photometric candidates with similar magnitudes in BOR99's survey still 
wait for membership confirmation indicate that the population of objects near 
0.015\,$M_{\odot}$ may be rather numerous in Orion. If the $\sigma$\,Orionis 
cluster IMF (B\'ejar et al. \cite{bejar00}) is representative of the field 
then old counterparts of these substellar bodies might be quite common in 
the galactic disk, and we would expect several tens populating the solar 
neighborhood ($d$\,$\le$\,10\,pc). At ages close to the Solar System their 
luminosities will have decreased by 4 orders of magnitude becoming extremely 
faint at all wavelengths, and their atmospheres will have cooled down to 
temperatures similar to those of Jupiter and Saturn, so it is expected 
their energy distributions resemble those of these planets. Objects with 
$\sim$0.01\,$M_{\odot}$ have been found at very close orbits around stars 
(Marcy \& Butler \cite{marcy98}, and references therein); their likely 
formation in a protoplanetary disk could be related to mass accretion onto 
a rocky core or to a gravitational collapse, or both. We cannot discard 
any of them as a possible origin for S\,Ori\,47, which may have been born 
orbiting a more massive cluster member. Because of its low mass S\,Ori\,47 
might have experienced severe orbital disruptions which ejected it from its 
birth place. However, given the young age of the $\sigma$\,Orionis cluster 
and its density ($n\,\sim\,12$\,objects\,pc$^{-3}$), this is not 
very likely because the number of encounters among the cluster stellar 
population is expected to be rather small (Laughlin \& Adams 
\cite{laughlin98}; Fuente Marcos \& Fuente Marcos \cite{fuente99}). 
S\,Ori\,47 could well have originated as a free-floating substellar 
cluster member, thus representing the natural extension of the IMF of 
objects formed via the process of cloud fragmentation.

\acknowledgments
We are indebted to I. Baraffe and the Lyon group for providing theoretical 
evolutionary trac\-ks prior to publication. We also thank A. Burrows for 
providing an electronic version of his models. We are grateful to D. 
Ardila for his cooperation in obtaining some of the data. Data presented 
here were obtained at the W. M. Keck Observatory, which is operated as a 
scientific partnership among CALTECH, the University of California and 
NASA; and at the TCS telescope 
operated on the island of Tenerife in the Spanish Observatorio del Teide 
of the IAC. Partial finantial support was provided by 
the Spanish DGES project PB95--1132--C02--01. E.L.M. was partially 
supported by a fellowship of the Spanish Ministry of Education. G.B. 
acknowledges the support of NSF through grant AST96--18439.

\clearpage

\plotone{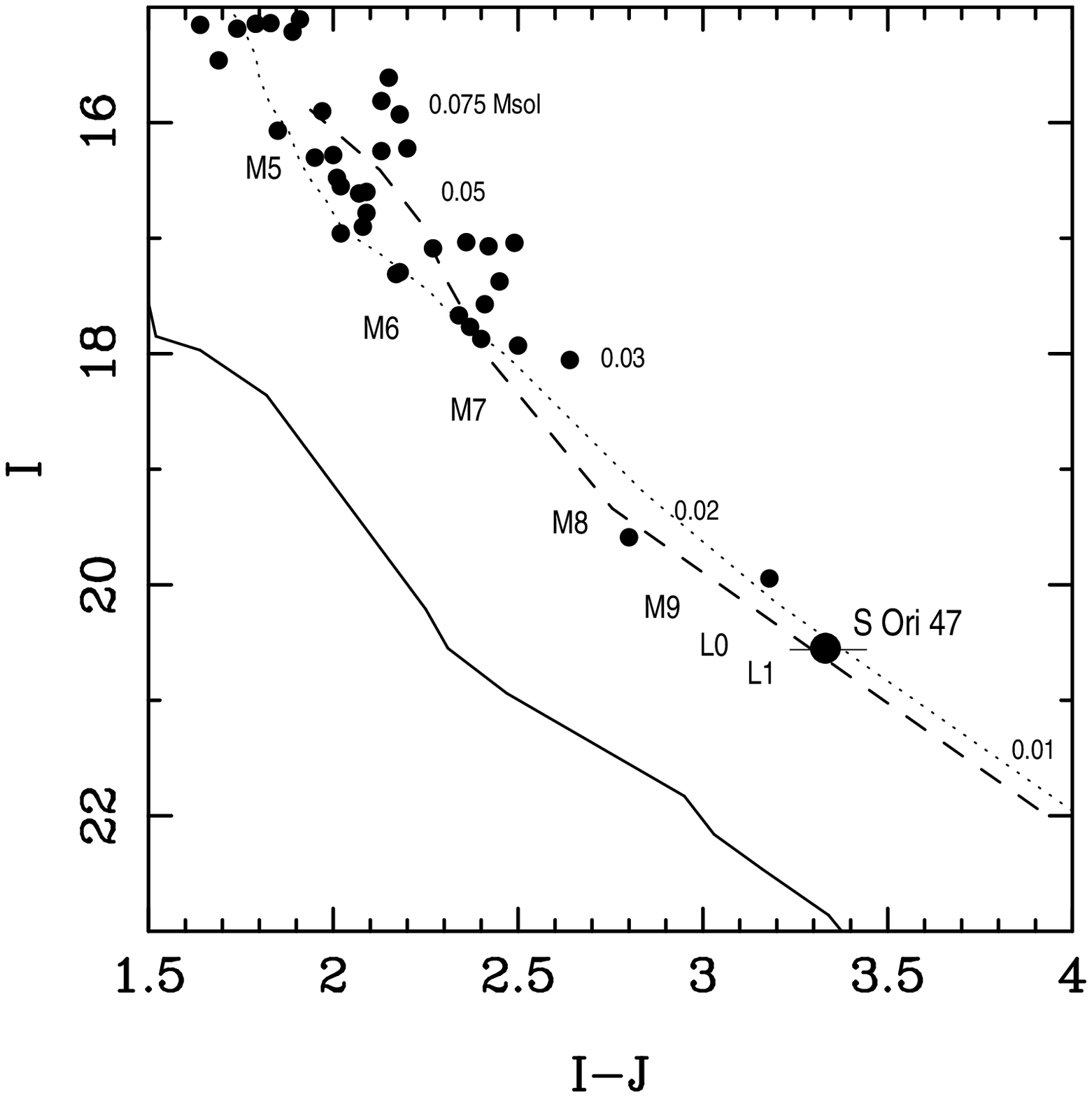}
\figcaption[osorio1.eps]{\label{fig1} 
Location of S\,Ori\,47 (large-size circle) in the optical-IR color-magnitude 
diagram. Some other $\sigma$\,Orionis substellar cluster members (small-size 
circles, B\'ejar et al. \cite{bejar00}) are also plotted. The solid line 
corresponds to the average sequence of very low-mass dwarfs shifted 
to the cluster distance. Theoretical isochrones of Burrows et al. (1997, 
3\,Myr, dashed line) and Baraffe et al. (1998, NextGen 5\,Myr, dotted line) 
are oveplotted to the data. Masses are given in solar units and spectral 
types are labeled as a function of the $I-J$ color.}

\plotone{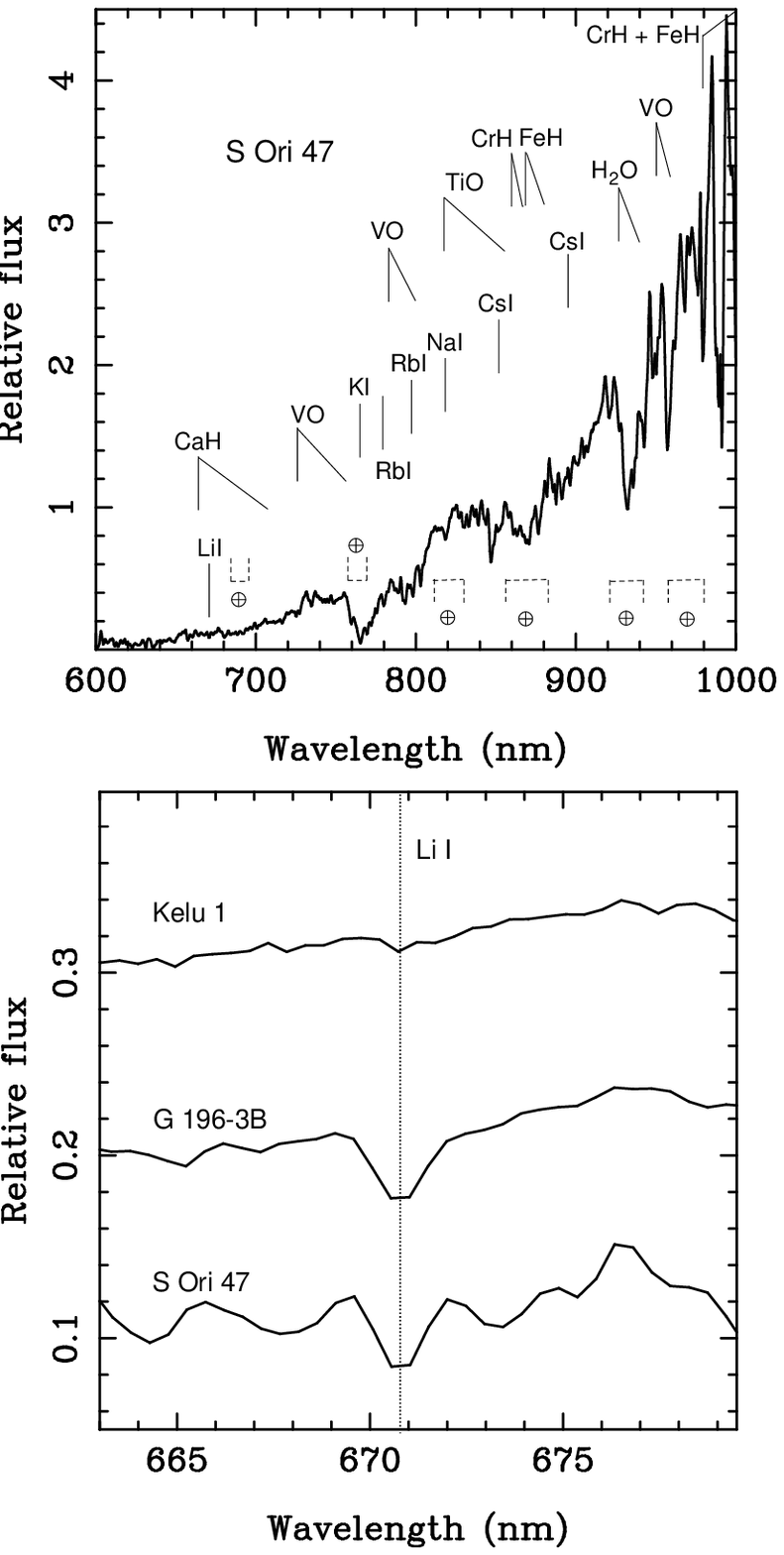}
\figcaption[osorio2.eps]{\label{fig2}
{\sl (Upper panel)} Keck low-resolution spectrum of S\,Ori\,47. It has 
been normalized at around 833.5\,nm. A boxcar smoothing of 3 pixels has 
been applied. Identification of some atomic and molecular features is given 
in the top. Telluric bands are marked with dashed lines. \ \ {\sl (Lower 
panel)} Enlargement of S\,Ori\,47's optical spectrum around the Li\,{\sc i} 
resonance doublet. Two other L1--L2 brown dwarfs in the field observed with 
identical instrumental configuration (displaced by 0.1 units each) are shown 
for comparison. }

\plotone{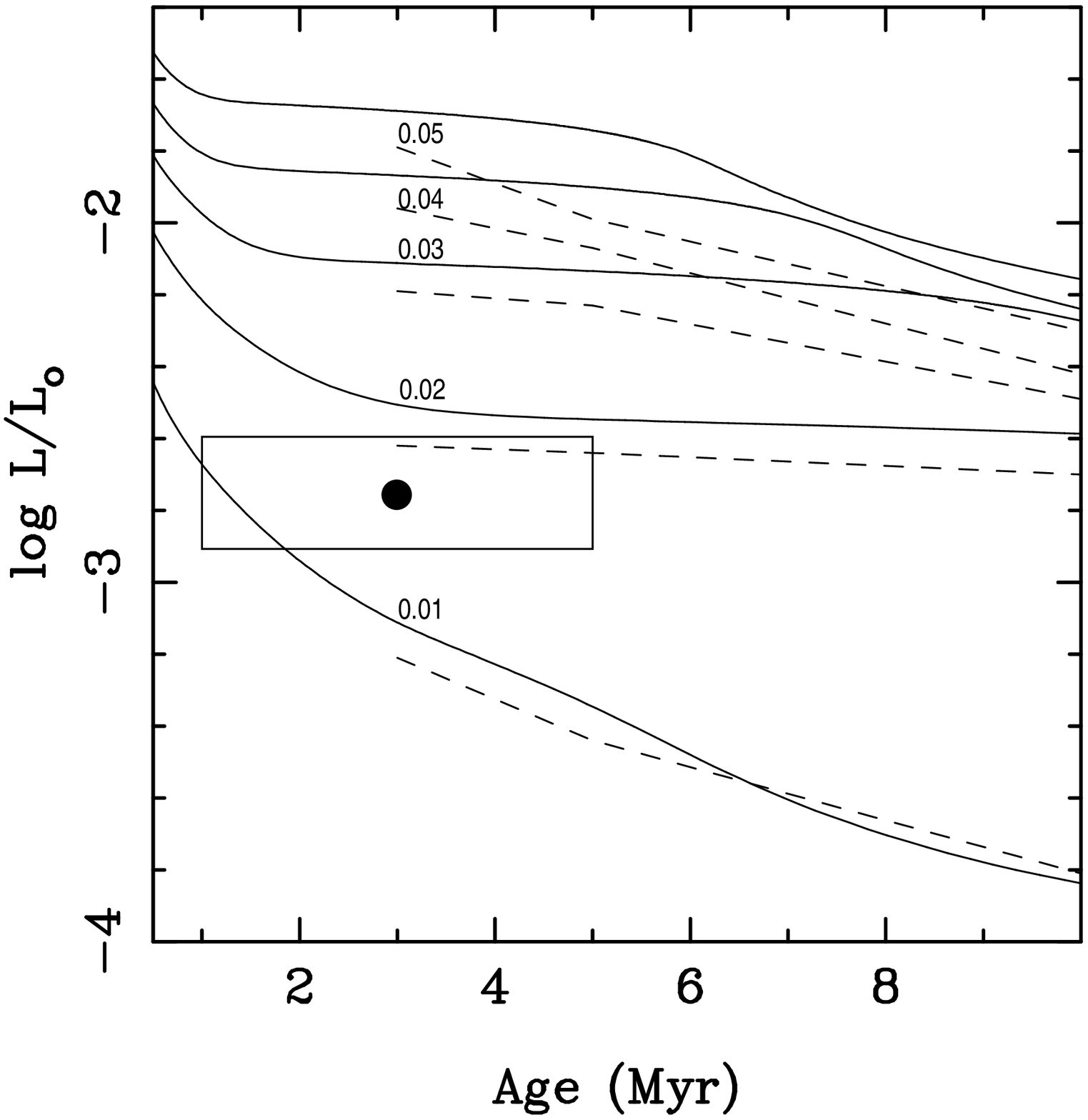}
\figcaption[osorio3.eps]{\label{fig3}
Time dependence of luminosity for very low-masses according to the 
theoretical models by Burrows et al. (\cite{burrows97}, solid line) and the 
Lyon group (1999, dashed line). The region occupied by S\,Ori\,47 
considering the uncertainties in the determination of its luminosity and 
age is indicated by the box. Masses are labeled in solar units. }

\clearpage

\begin{deluxetable}{ccccccccccc}
\scriptsize
\tablecaption{Observed data and basic parameters for S\,Ori\,47. \label{tab1}}
\tablewidth{0pt}
\tablehead{
 \colhead{$R-I$\tablenotemark{a}} & \colhead{$I-Z$\tablenotemark{a}} & 
 \colhead{$I-J$} & \colhead{$J_{\rm UKIRT}$} & 
 \colhead{$T_{\rm eff}$\tablenotemark{b}} & 
 \colhead{log $L/L_{\odot}$} & \colhead{Sp. Type} & \colhead{H$\alpha$} & 
 \colhead{Li\,{\sc i}$_{670.8}$} & \colhead{Na\,{\sc i}$_{819.5}$} & 
 \colhead{Cs\,{\sc i}$_{852.1}$}
}
\startdata
2.4$\pm$0.3 & 1.00$\pm$0.06 & 3.3$\pm$0.1 & 17.2$\pm$0.1 & 2000$\pm$100\,K & 
--2.75$\pm$0.15 & L1.5$\pm$0.5 & $\le$6\tablenotemark{c} & 
4.3$\pm$0.5\tablenotemark{c} & 2.5$\pm$0.5\tablenotemark{c} & 
1.2$\pm$0.5\tablenotemark{c} \nl
\enddata
\tablenotetext{a}{Colors taken from B\'ejar et al. (\cite{bejar99}).}
\tablenotetext{b}{Assumes a dwarf spectral type conversion.}
\tablenotetext{c}{Equivalent widths are given in \AA, wavelenghts in nm.}
\end{deluxetable}



\clearpage
 
\begin{thebibliography}{}
\bibitem[1998]{baraffe98} Baraffe, I., Chabrier, G., Allard, F., \& 
         Hauschildt, P. 1998, \aap, 337, 403

\bibitem[1996]{basri96} Basri, G., Marcy, G. W., \& Graham, J. R. 1996,
         \apj, 458, 600

\bibitem[1999]{basri99} Basri, G., et al. 1999, \apj, submitted

\bibitem[1999]{bejar99} B\'ejar, V. J. S., Zapatero Osorio, M. R., \& 
         Rebolo, R. 1999, \apj, 521, 671 (BOR99)

\bibitem[2000]{bejar00} B\'ejar, V. J. S., et al. 2000, in preparation

\bibitem[1998]{bouvier98} Bouvier, J., Stauffer, J. R., Mart\'\i n, E. L.,
         Barrado y Navascu\'es, D., Wallace, B., \& B\'ejar, V. J. S. 1998,
         \aap, 336, 490

\bibitem[1997]{burrows97} Burrows, A., Marley, M., Hubbard, W. B., et al. 
         1997, \apj, 491, 856

\bibitem[1992]{casali92} Casali, M. M., \& Hawarden, T. G. 1992, JCMT-UKIRT 
         Newslt., 4, 33

\bibitem[1998]{comeron98} Comeron, F., Rieke, G. H., Claes, P., Torra, J., 
         \& Laureijs, R. J. 1998, \aap, 335, 522

\bibitem[1999]{comeron99} Comeron, F., Rieke, G. H., \& Neuh\"auser, R. 
         1999, \aap, 343, 477

\bibitem[1994]{dantona94} D'Antona, F., \& Mazzitelli, I. 1994, \apjs, 90, 467

\bibitem[1997]{dantona97} D'Antona, F., \& Mazzitelli, I. 1997, Mem.
         Soc. Astron. Italiana, 68, 807

\bibitem[1997]{delfosse97} Delfosse, X., Tinney, C. G., Forveille, T.,  
         et al. 1997, \aap, 327, L25

\bibitem[1999]{delfosse99} Delfosse, X., Tinney, C. G., Forveille, T., 
         et al. 1999, \aaps, 135, 41

\bibitem[1999]{fuente99} Fuente Marcos, C., \& Fuente Marcos, R. 1999, 
         New. Astron., 4, 21

\bibitem[1999]{gizis99} Gizis, J. E., Reid, I. N., \& Monet, D. G. 1999, 
         \aj, in press

\bibitem[1999]{kirk99} Kirkpatrick, J. D., Reid, I. N., Liebert, J., 
         et al. 1999, \apj, 519, 802

\bibitem[1998]{laughlin98} Laughlin, G., \& Adams, F. C. 1998, \apj, 508, L171


\bibitem[1998]{leggett98} Leggett, S. K., Allard, F., \& Hauschildt, P. H. 
         1998, \apj, 509, 836


\bibitem[1999]{luhman99} Luhman, K. 1999, \apj, in press

\bibitem[1997]{luhman97} Luhman, K., Liebert, J., \& Rieke, G. H. 1997,
         \apj, 489, L165

\bibitem[1998]{luhman98} Luhman, K., Rieke, G. H., Lada, C. J., \& 
         Lada, E. A. 1998, \apj, 508, 347

\bibitem[1998]{magazzu98} Magazz\`u, A., Rebolo, R., Zapatero Osorio, M. R., 
         Mart\'\i n, E. L., \& Hodgkin, S. T. 1998, \apj, 497, L47

\bibitem[1998]{marcy98} Marcy, G. W., \& Butler, R. P. 1998, ARA\&A, 36, 57

\bibitem[1997]{martin97} Mart\'\i n, E. L., Basri, G., Delfosse, X., 
         \& Forveille, T. 1997, \aap, 327, L29

\bibitem[1999a]{martin99a} Mart\'\i n, E. L., Basri, G., \& Zapatero Osorio, 
         M. R. 1999a, \aj, in press

\bibitem[1998]{martin98} Mart\'\i n, E. L., Basri, G., Zapatero Osorio, M.
         R., Rebolo, R., \& Garc\'\i a L\'opez, R. J. 1998, \apj, 507, L41

\bibitem[1999b]{martin99b} Mart\'\i n, E. L., Delfosse, X., Basri, G., 
         Goldman, B., Forveille, T., \& Zapatero Osorio, M. R. 1999b,
         \aj, in press

\bibitem[1996]{martin96} Mart\'\i n, E. L., Rebolo, R., \& Zapatero Osorio, 
         M. R. 1996, \apj, 469, 706

\bibitem[1995]{oke95} Oke, J. B., et al. 1995, \pasp, 107, 375

\bibitem[1999]{oppenheimer99} Oppenheimer, B. R., Kulkarni, S. R., \& 
         Stauffer, J. R. 1999, in Protostars and Planets IV, V. Mannings, 
         A. Boss, S. Russell, eds. (Tucson: University of Arizona Press), 
         in press

\bibitem[1999]{pavlenko99} Pavlenko, Ya., Zapatero Osorio, M. R., \& 
         Rebolo, R. 1999, \aap, submitted

\bibitem[1998]{rebolo98} Rebolo, R., Zapatero Osorio, M. R., Madruga, S., 
         B\'ejar, V. J. S., Arribas, S., \& Licandro, J. 1998, Science, 
         282, 1309

\bibitem[1995]{rebolo95} Rebolo, R., Zapatero Osorio, M. R., \& Mart\'\i n, 
         E. L. 1995, \nat, 377, 129

\bibitem[1999]{reid99} Reid, I. N., Kirkpatrick, J. D., Liebert, J., 
         et al. 1999, \apj, 521, 613

\bibitem[1997]{ruiz97} Ruiz, M. T., Leggett, S. K., \& Allard, F. 1997, 
         \apjl, 491, L107

\bibitem[1996]{saumon96} Saumon, D, Hubbard, W. B., Burrows, A., Guillot, T., 
         Lunine, J. I., \& Chabrier, G. 1996, \apj, 460, 993

\bibitem[1998]{stauffer98} Stauffer, J. R., Schild, R., Barrado y 
         Navascu\'es, D., et al. 1998, \apj, 504, 805

\bibitem[1997]{tinney97} Tinney, C. G., Delfosse, X., \& Forveille, T., 
         1997, \apj, 490, L95

\bibitem[1997]{walter97} Walter, F. M., Wolk, S. J., Freyberg, M., \& 
         Schmitt, J. H. M. M. 1997, Mem. Soc. Astron. Italiana, 68, 1081

\bibitem[1997]{osorio97} Zapatero Osorio, M. R., Rebolo, R., Mart\'\i n, 
         E. L., et al. 1997, \apj, 491, L81

\bibitem[1999]{osorio99} Zapatero Osorio, M. R., Rebolo, R., Mart\'\i n, 
         E. L., Hodgkin, S. T., Cossburn, M. R., Magazz\`u, A., 
         Steele, I. A., \& Jameson, R. F. 1999, \aaps, 134, 537

\end{thebibliography}
\end{document}